# A Hybrid LTR-based System via Social Context Embedding for Recommending Solutions of Software Bugs in Developer Communities

**Fouzi Harrag[1]\*, Mokdad Khemliche[2],**

[1,2]Computer Sciences Department,
College of Sciences,
Ferhat Abbas University,
Setif, Algeria,
[1]fouzi.harrag@univ-setif.dz, [2]khemliche.mokdad@univ-setif.dz

\* Corresponding author: Fouzi Harrag (fouzi.harrag@univ-setif.dz).

*Abstract* – Questions and Answering forums such as Stack Overflow play an important role in supporting software developers in finding answers to queries related to issues such as software errors and bugs. However, searching through a large set of candidate answers could be time consuming and may not lead to the best solution. In this research, the effectiveness of data mining models and machine learning techniques to solve this kind of problems is evaluated. We propose a recommender system to aid developers in finding solutions to their software bugs by carefully mining Stack Overflow. The proposed model leverages the knowledge available through crowdsourcing the Q&A available in Stack Overflow to recommend a solution to software bugs. We use deep learning techniques to construct the required Learning-to-Rank (LTR)-based model using the social context embedding the Stack Overflow features. Text mining, natural language processing and recommendation algorithms are used to extract, evaluate and recommend the best relevant bug solutions. Additionally, our model achieves nearly 78% correct solutions when recommending the 10 best answers for each question.

*Keywords* – Recommender System, learning to rank, Mining software repositories, Text Mining, Deep learning, Stack Overflow.

## 1. INTRODUCTION

In the software development area, there is huge amount of unstructured data that grows fastly every day. This data exists at different levels and produced by many systems that are used in the software development process such as versioning systems, issue trackers, design documentation and many others. Mining the rich software engineering data has attracted much interest lately in the data mining domain opened new research and developments directions. It is expected that the results of this research will contribute to the development and maintenances activities of the software development process.

### 1.1. Software data

Unstructured data refer to information that is not organized by following a precise schema or structure. Such data often include text (e.g., email messages, software documentation, and code comments) and multimedia (e.g., video tutorials, presentations) contents. These kinds of data are estimated to represent 80% of the overall information created and used by enterprises in software projects [1]. Large number of artefacts are generated in the development process. This huge data is continuously growing over time and is known as and organized in software repositories.

### 1.2. Software repositories

Ahmed E. Hassan define software repositories [1] as a record-keeping database that stores data about artifacts of a complex computer-based system. It tracks changes applied to the artifact and stores corresponding Meta data. Source control repositories, bug repositories, archived communications, deployment logs, and code repositories are examples of software repositories that are commonly available for most software projects.

### 1.3. Stack Overflow as a Knowledge Repository

Social media, unlike traditional media, gives people an easy way to communicate, collaborate and share information with each other. There are many types of social media such as blogs, micro blogs, bookmarking sites, social news, media sharing and social networks. Nowadays, social media are turning into social productivity through the participation of individuals discussing their ideas, experiences and skills to generate content. This collective work is called crowd sourcing. Brayvold [2] defined crowd sourcing as "the process of getting work or funding, usually online, from a crowd of people". It generates content by the participation of a large group of people contributing with their skills, ideas, and knowledge. The usage of crowd sourcing can help companies and individual in performing their tasks at a low cost compared to employing specialists. Furthermore, one has access to a high number of people with expertise and working at different times. One of the most popular crowd sourcing platforms for software development is stack overflow.

Stack overflow is a public platform with a large collection and knowledge base for computer scientists, developers and programmers and is crow sourced. However, extracting and finding the right information for a specific problem can be challenging. Hence, using data mining to develop a tool to help users in finding the required information will be of great interest to the community using Stack Overflow. In section 2, we will outline and clarify the concept of mining software repositories.

### 1.4. Mining Software repositories

The field of mining software repositories (MSR) aims at examining and analyzing "the rich data available in software repositories to uncover interesting and actionable information about software projects and systems" [3].

### 1.5. Research contributions

The aim of our work is to design and develop a recommender system that will help Stack Overflow users in finding the best answers to their queries to solve their software bugs problems using data mining. We are proposing a deep learning-based model that will be able to rank the best relevant solutions for any software bug query as this is a ranking problem. Based on our initial study, this work makes the following new contributions in the research area of bug solutions recommendation systems.

- We clearly demonstrate through our experiences in this paper that Stack Overflow as a question-and-answer program translates a wealth of valuable knowledge about the experiences of participating developers. This huge amount of useful information represented by the responses and comments of developers and programmers and their interactions within the Stack Overflow system is being exploited in the process of recommending bug solutions.
- In our proposed model, a novel bug solutions recommendation approach based on the incorporation of the content with the social interaction's context on Stack Overflow into a unified learning-to-rank schema is proposed.
- The learning of our LTR ranking model is based on the extraction of four different kinds of features i.e., statistical, textual, context and popularity features, to enable the ranking of the solutions in real-time for recommendation.
- In terms of technologies, we proposed a technique for the automated recommendation of relevant Stack Overflow posts to developers and programmers to support them in maintaining and fixing the bugs in their code and applications.
- In terms of tools, we implemented our proposed technique by designing a prototype along with proposing two types of model baseline variations for the purpose of evaluation.
- In terms of evaluation, we have evaluated our proposed model and its two baseline variants on a benchmark of 29395 queries and answers extracted from Stack overflow. Our proposed system achieves better results and makes progress compared to the baselines.
- In terms of user experience study, a small-scale user study is presented, with the use of 2 evaluators for the qualitative evaluation of the performance of our system. The user study is based on comparing the results recommended by our proposed system with Google search and Stack Overflow search. The results clearly show that our proposed system outperforms Google and Stack Overflow search engines in the specific task of recommending bug solutions.

This article is organized as follows: In Section 2, we summarize the state of the art in the field of mining software repositories and give the definitions of data mining, recommender systems and NLP techniques. In Section 3, we present an overview of related work in the field of mining software repositories with a discussion of the different proposed approaches. Section 4 is reserved for the presentation of the approach proposed for the development of our system. In this section, we focus more on the design aspects while taking into account the details related to the integration of machine and deep learning techniques in the construction of our framework. Section 5 will be devoted to the presentation of the implementation details of our system while the whole discussion of the results obtained is presented in Section 6. Section 7 will conclude our study and suggests some future research directions.

## 2. Related Works

Many researchers have discussed the effectiveness of using data mining techniques to facilitate the debugging process for software engineering developers. This section presents an overview of the state of the art of this research area.

Xin et al [4] presented a ranking approach that simulates the bug locating process used by developers. The ranking model benefited from domain knowledge such as API specifications, the bug-fix history, and code-change history. The ranking score of each source file is calculated as a weighted combination of an array of features. The evaluation of the experimental results was performed on six open sources java projects, which are Eclipse, JDT, Birt, SWT. Tomcat and AspectJ. The results showed that their ranking approach is better than VSM-based BugLocator [Ref1] and LDA-based BugScout [Ref2] approaches. Their method assigns the relevant files to over 70% of the bug reports within the top 10 recommendations for Eclipse and Tomcat projects.

Rafi et al. [5] proposed an automated approach for finding and ranking potential relevant classes for bug reports. Their approach used a multi-objective optimization algorithm to find balance between minimizing the number of recommended classes and maximizing the correctness of the proposed solution. Based on the use of the history of changes and bug-fixing, and the lexical similarity between the bug report description and the API documentation estimated the correctness of the recommended classes. They evaluated their system on six large open-source Java projects. The experimental results showed that the search-based approach was better than mono-objective formulations Lexical-based Similarity (LS) and History-based Similarity (HS). Their search-based approach can find the true buggy methods for over 87% of the bug reports within the top 10 recommendations.

Lam et al., [6] presented an integrated approach between deep neural network (DNN) and Revised Vector Space Model (rVSM), and an information retrieval (IR) technique to locate and rank potential relevant classes for bug reports. rVSM gathers the textual similarity feature between bug reports and source files. DNN is used to learn to relate the terms in bug reports to potentially different code tokens and terms in source files. The Evaluation of their approach was on real-world bug reports in open-source projects. Combining DNN with their new model achieved high accuracy of bug localization than the state-of-art IR and machine learning techniques.

The approach that was used to benefit from the "crowd knowledge" available in stack overflow to aid developers in their activities was presented in [7]. This strategy recommended a ranked list of question-answer pairs from stack overflow based on a query. The ranking criteria was based on the textual similarity of the pairs with respect to query the quality of the pairs, and a filtering mechanism that considers only "how-to" posts. using a *Lucene + Score + How-to* approach that uses a combination of the Vector Space Model (VSM) of Information Retrieval and the Boolean model to determine how relevant a given Document is to a User's query. In this paper, the researchers an experiment about programming problems on three different topics (Swing, Boost and LINQ) which are frequently used by the software development community. By analyzing the results, it was found that the *Lucene + Score + How-to* approach proved to be useful in solving programming problems by 77.14% when used on the assessed activities that have at least one recommended pair.

Fabio et al. [8] discussed the question of why developers are abandoning from legacy developer forums to stack overflow platform making a lot of crowd sourced knowledge at risk of being left behind. They aimed to add to the body of evidence of existing research on best-answer predication. They performed an experiment using data from Stack Overflow to train a binary classifier. Then, they tested a classifier on a dataset retrieved from the legacy document using support forum. The findings showed that their model could find best answers with a good accuracy when all features are enabled e.g. answer up votes, number of sentences and answer length. The results gave a positive proof towards the automatic migration of crowd-sourced knowledge from legacy forums to modern Q&A sites.

Perricone [9] utilizes the network structure of Stack Overflow to recommend a set of related questions for a given input question. In particular, the project employs a modified guided Personalized Page Rank algorithm to generate candidate recommendations and compares the results to those recommended by Stack Overflow. Semantic similarity and tag-overlap were used to assess candidate recommendations. For a given recommendation set, the average text, title, and tag-overlap scores were calculated. These scores were then averaged across all trials of the experiment to yield a final score.

Weld et al., [10], investigated the problem of systematically mining question-code pairs from Stack Overflow (in contrast to heuristically collecting them). They formulated the problem as a predicting problem as whether a code snippet is a standalone solution to a question. They proposed a novel Bi-View Hierarchical Neural Network that can capture both the programming content and the textual context of a code snippet (i.e., two views) to make a prediction. On two manually annotated datasets in Python and SQL domain, the framework substantially outperforms heuristic methods with at least 15% higher F1 and accuracy. Furthermore, they presented StaQC (Stack Overflow Question-Code pairs), the largest dataset to date of ~148K Python and ~120K SQL question-code pairs, automatically mined from Stack Overflow using this framework.

Beyer et al., [11] aim to automate the classification of Stack Overflow posts into seven question categories. As a first step, they have manually created a curated data set of 500 Stack Overflow posts, classified into the seven categories. Using this dataset, they applied machine learning algorithms (Random Forest and Support Vector Machines) to build a classification model for Stack Overflow questions. They then experimented with 82 different configurations regarding the preprocessing of the text and representation of the input data. The results of the best performing models show that their models can classify posts into the correct question category with an average precision and recall of 0.88 and 0.87 respectively when using Random Forest and the phrases indicating a question category as input data for the training. The obtained model can be used to aid developers in browsing Stack Overflow discussions or researchers in building recommenders based on Stack Overflow.

Zhang et al., [12] proposed a novel approach named Recommends Frequently Encountered Bugs (RFEB). By analyzing the content of Stack Overflow, RFEB try to recommends the frequently encountered bugs (FEBugs) that may affect many other developers. Among the plenty of questions posted in Stack Overflow, many of them provide the descriptions and solutions of different kinds of bugs. Unfortunately, the search engine that comes with Stack Overflow is not able to identify FEBugs well. To address the limitation of the search engine of Stack Overflow, they propose RFEB, which is an integrated and iterative approach that considers both relevance and popularity of Stack Overflow questions to identify FEBugs. To evaluate the performance of RFEB, they performed experiments on a dataset from Stack Overflow, which contains more than ten million posts. Finally, they compared this model with Stack Overflow's search engine on 10 domains, and the experiment results show that RFEB achieves the average $NDCG10$ score of 0.96, which improves Stack Overflow's search engine by 20%.

### *Discussion of related works*

We noted from the previous state of the art that most of the studies like [4] [5] and [6] provide a ranking approach that leverages domain knowledge to locate a bug by ranking all the source files likely to contain the cause of the bug. The researchers used the same benchmark datasets for evaluation in [4] [5] [6]. There is a similarity between [6] and [7], as both used a vector space model for ranking whereas [5] used Multi-objective algorithms called NSGA-II. [6] used Revised Vector Space Model (rVSM) and Deep Neural Network (DNN). Data mining techniques that were used in these studies are different such as [8] and [11] used classification, [9] used a modified version of page-rank, [10] used a novel Bi-View Hierarchical Neural Network algorithm and [12] used a novel RFEB approach to recommends frequently encountered bugs. The researchers suggested in [6] and [7] to improve their approaches by leveraging additional types of domain knowledge and using the SVM ranking with nonlinear kernels. They evaluated their approaches on different datasets of programming languages codes. In this paper, have taken into consideration these suggestions in the design and development of our proposed framework. Table 1 summarizes the reviewed studies.

Table 1: Summary of Related Works.

| Ref | Year | Method used | Data set | Results | Performances Evaluation |
|---|---|---|---|---|---|
| Xin et al., [4] | 2016 | - Ranking model<br>- VSM | Benchmark datasets from open source projects:<br>Eclipse ; JDT; Birt; SWT | The learning rank approach achieved better results than the Bug Locator, VSM, Usual suspects on all six projects. | 1) Eclipse: accuracy =80%, MAP= 0.44 ,MAR=0.51 .<br>2) JDT: accuracy =80%, MAP=0.39, MRR=0.51 .<br>3) Birt: accuracy=50%, MAP=0.16, MRR=0.21. |
| Rafi et al., [5] | 2016 | - Multi-objective algorithm (NSGA-II)<br>- Rearch-based algorithms | Benchmark datasets form open-source systems :<br>EclipseIU; Tomcat; AspectJ<br>Birt; SWT; JDT. | On all the 6 systems, the NSGA-II algorithm shows better results by comparing it with the random search and the mono objective formulations of similarity based on the lexicon (LS), similarity based on the history (HS) and ( GA). | 1) EclipseIU : precision=82%,recall=79%, accuracy=80%.<br>2) Tomcat: precision=91%,recall=81%, accuracy=90%.<br>3) AspectJ: precision=79%, recall=86%, accuracy =88%. |
| Lam et al., [6] | 2017 | Revised Vector Space Model (rVSM) + Deep Neural Network (DNN) | Benchmark datasets from open-source projects: Aspectj ,Birt ,Eclipe platform ,JDT,SWT,Tomcat . | DNNLOC achieves highest accuracy with the combination of relevancy via DNN, textual similarity via rVSM, and the metadata features. | 1) TomCat :accuracy=80.4%,MRR=0.60,MAP=0.52<br>2) AspectJ: accuracy=85%, MRR=0.52, MAP=0.32 |
| Eduardo et al., [7] | 2016 | Logistic regression classifier (LR)+ Normalized Discounted Cumulative Gain (NDCG) | Dataset from Stack Overflow (March 2013 version) | The Lucene+ Score+ Howto approach achieved better performance than Google on Boost. | $NDCG_{Relev}$=0.3583<br>$NDCG_{Reprod}$=0.5243 |
| Fabio et al., [8] | 2016 | Alternating Decision Trees (ADT) classifier + information gain (IG) | 1). Dataset from Docusign , it is a legacy forum.<br>2) Dataset from Stack Overflow | The model can find best answers with a good performance, when all features are enabled | Accuracy ~90%,<br>F=.86, AUC=.71. |

## 3. Proposed Approach

Our framework is using the concept of recommender systems to model the problem of mining stack overflow in order to find the Top-k solutions for a bug report.

### 3.1. Overall Framework

By looking to the architecture of our model, we find that a bug report is issued as input and the ranked list of related best answers is recommended as output. First, when a new bug report is received, the preprocessing steps are started. The Information extracted from the bug report are then formulated as query and issued to the index of questions. A similarity measure between the query and a set of answers is calculated. The N best selected answers are then passed to the kernel of our model. The list of proposed solutions is then re-ranked by score. The learning to rank approach is applied to train a ranking model that use many features extracted from the Stack Overflow dataset. Figure 1 shows the overall architecture of our system framework.

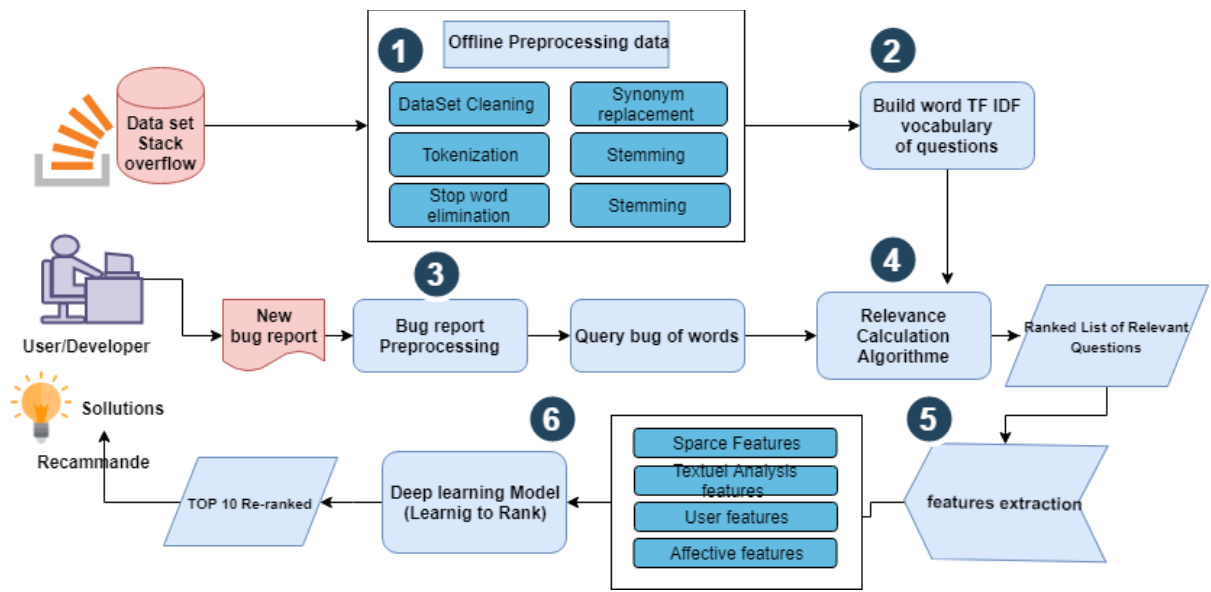

*Figure 1:* The overall architecture of our system framework

### 3.2. Stack Overflow Dataset description

The Stack Overflow team periodically publishes updated data dumps under a Creative Commons cc-wiki license that allow the use of the data for the purpose of analysis. In this study we used the version published on March 4, 2019, which is the most up-to date at the time this work began. In their compressed form, the size of these files is approximately 70 gigabytes.

Stack Overflow dataset content is stored as follows:

• Each post, whether it is a question or an answer, is stored in a Post file.

• An idParent is used to link each question with its related answer posts.

• All posts are subject to the users voting.

- The post edits history is kept in a file completely separate from the current content version.

- There is also a separate file called Comments that is used to store users' comments on questions or answers.

Table 1 below shows how the dataset is separated into 8 XML files after decompressing the files.

| File name | File Description |
|---|---|
| Badges.xml | Stack Overflow User Badges |
| Comments.xml | Comments present in Stack Overflow Posts |
| PostHistory.xml | Stack Overflow Revision information from Posts |
| PostLinks.xml | Stack Overflow Related and/or duplicated post links |
| Posts.xml | Actual posts or current version of Stack Overflow |
| Tags.xml | Stack Overflow Lists of Tags |
| Users.xml | Stack Overflow Users Information |
| Votes.xml | Stack Overflow Votes for Posts (Q/A) |

*Table 1* The structure of Stack Overflow Data Dump

In our work, we focus on Posts, Comments and Users files. Table 2 shows the most important part of these data where each attribute is described briefly.

| Id | Unique identifier for each question or answer |
|---|---|
| PostTypeId | 1 = Question, 2 = Answer |
| AcceptedAnswerId | for a question, it represents the Id of the accepted answer |
| ParentId | Id of the question an answer is associated with |
| CreationDate | Date time of the post creation |
| Score | Number of Upvotes - Downvotes for a post |
| ViewCount | Times the post was viewed |
| Body | Text of the question or answer (HTML) |
| OwnerUserId | User Id of the post |
| LastEditorUserId | User Id of the last editor of the post |
| LastEditorDisplayName | User display name of the last editor of the post |
| LastEditDate | Datetime of the most recent edit to the post |
| LastActivityDate | Datetime of the last action on the post |
| Title | Title of a question (null if answer) |
| Tags | Associated tags of the question, e.g., Java, Android, |
| AnswerCount | Number of answers for the question (null if no answers) |
| CommentCount | Number of comments on post |
| FavoriteCount | Number of times the post has been favorite |
| ClosedDate | Datetime when the post was closed (null if the post is open) |
| CommunityOwnedDate | Datetime when the post was community wikied |

*Table 2* Schema of the Posts dataset

### 3.3. Storing Data in PostgresSQL database

After the completion of the downloading operation, the zip file was decompressed into 8 files. The Posts file was about 66 gigabytes, which makes it difficult to handle a file of this large size. A Python script was developed to correctly parse the three files (posts, comments, users) and insert the extracted data into a database. This process took approximately 65 minutes in total to finish. Once the data has been loaded into the table, SQL queries can be used to manage the data in an extremely powerful and flexible way.

### 3.4. Preprocessing of posts and bugs

After storing our data into a PostgreSQL[1] database, we made it easy to manipulate it by executing many queries to select specific needed data. In this Stage, We implemented the different steps of the Natural Language Processing using the python *NLTK*[2] package, which is a tool Kit dedicated for operations related to Natural language. We can summarize our work into the following steps:

- ✓ Dataset cleaning: each post's body contains two main parts: plain text and source code. Special tags such as "<code>" and "<pre>" are used to separate these two parts.
- ✓ We removed this this tags using *beautifulSoup*[3] that is a python package for parsing HTML and XML document.
- ✓ We Concatenated the Title, normal text and all code segments.
- ✓ We created the function *preprossing_Text* that take a String and return clean tokens. This function is used in:
  - Splitting the textual data passed in *arg* using *word_tokenize()* function.
  - The punctuation library is used to remove punctuation marks, extra spaces and facilitate the Trimming process.
  - Removing all the Stop-Words English.
  - Stemming Tokens using *Porter* Stemmer algorithm of *NLTK* package.

For the query preprocessing (bug report in our context), we concatenated and merged all the textual information such as the title, description and steps to reproduce them as unique String. We applied the same preprocessing functions already applied on the database to this new generated string representing the data extracted from the bug.

### 3.5. Building TF-IDF Index from the questions vocabulary

In this stage, we created the TFIDF index from the questions vocabulary. This index is used as an input layer to our recommender system for selecting the most important questions. We made a first query to select All questions (this means selecting all posts with *postTypeId*=1) and exported them into a csv file. This query returns all unedited questions from previous years, ordered by RANDOM. After completing all preprocessing Steps, all questions are then parsed into tokens (features). For each feature, we calculate a $TFIDF$ metric using *TfidfVectorizer* class of *sklearn* module. This class is used for the conversion of our collection of raw documents into a matrix of TF-IDF features.

### 3.6. Extracting and preparing Features

Our proposed approach aim to build a ranking estimator based on heterogeneous dense and sparse textual features using deep learning techniques. The stage of features selection play an important role in this approach by selecting the most useful features for our model. The neural network model is built based on a large percent of embedding/sparse features of question answer pair of *Stack Overflow* dataset. In order to extract these features, we created another *SQL* query to select a large portion of data by self-joining the posts. The preprocessing step helped us to extract a set of embedding features from the question-answer pairs in order to create the global vocabulary. In addition, we also used the metadata/dense features from posts and users with some other extra textual features for the improvement of the performance of our model. A new python class *PostTextAnalysis* is created for extracting textual features from the body and title sections of each post. Table 3 shows the details for each feature.

---
[1] https://www.postgresql.org/

[2] https://www.nltk.org/

[3] https://www.crummy.com/software/BeautifulSoup/bs4/doc/

| Feature | Value |
|---|---|
| Body Length | Count of characters of the post (HTML) |
| Email Count | Count of e-mail addresses in post |
| URL Count | Count of URLs in the post |
| Uppercase | Percentage of uppercase characters |
| Lowercase | Percentage of lowercase characters |
| Spaces Count | Count of spaces |
| ARI | $4.71 * \frac{characters}{words} + 0.5 * \frac{words}{sentences} - 21.43$ |
| Flesch Reading Ease | $206.835 \left(1.015 * \frac{words}{sentences}\right)\left(84.6 * \frac{syllables}{words}\right)$ |
| Flesch-Kincaid Grade | $\left(0.39 * \frac{words}{sentences}\right) + \left(11.8 * \frac{syllables}{words}\right) - 15.59$ |
| Gunning Fog Index | $0.4 \left(\frac{words}{sentences}\right) + prcentageComplexWord)$ |
| SMOG Index | $\sqrt{\left(complexwords * \frac{30}{sentences}\right)} + 3$ |
| Coleman Liau Index | $\left(0.588 * \frac{characters}{words}\right) - \left(0.296 * \frac{words}{sentences}\right) - 15.8$ |
| LIX | $\frac{words}{periods} + \frac{longwords * 100}{words}$ |
| RIX | $\frac{longwords}{sentences}$ |
| Sentiment | Polarity [-1.0, 1.0], [Negative, Positive] |
| Subjectivity | [0.0, 1.0], [Objective, Subjective] |
| Lines of Code | Count of lines between tags <code> |
| Code Percentage | Percentage of a posts lines that are code |
| Number of Code | Count of <code> tag |
| Number of P Tags | Count of <p> tags |
| Title-Body Similarity | Similarity between the body and the question's title |
| Title Length | Number of characters in the title of any question |

*Table 3 Features extracted using PostTextAnalysis*

*LABEL_FEATURE (The answer relevance label)*

This feature is used to represent the relevance judgments of query and document pair in ranking problem. This label can be explicit (human ratings) or implicit (clicks). This is representing the class that will be predicted in the learning model. The *Label_Feature* can be relevant or not relevant or a simple value representing the grade of relevance. In our case, the score attribute of answer stored in the post table is used as Label_Feature. The stack overflow platform provide a mechanism of voting for answers. Users vote for answers as good or bad and the vote scoring is updated by calculating the difference between the Up-Vote and Down-Vote. Many answers scores have a high values (thousands) while others have a low values. Using all this classes in the deep learning model is not appropriate. So, to reduce this problem, we created partitions using the *Numpy* library. For each question, the relevance list of answers will be divided into 5 interval (list) whose relevance are graded on a scale of 1-5. Figure 2 shows an example of this grading operation.

```
[[1, 2, 2, 2], [11, 16, 17, 22], [24, 30, 32, 36], [45, 69, 79, 125], [361, 943, 1860]]
Score :
30
grade of relevance :
3
```

*Figure 2* *an example of relevance lists*

The number {1, 2, 11, 16, 943, 1680…etc.} represent votes scoring of the answers that are given by users in a stack overflow post. Using the *Numpy* library, we tried to create 5 partitions grouped by order of score from low to high-grade value of relevance. For getting the grade of relevance of the answer that have score 30, we should find the group of this score. In this case, the score 30 belongs to the group 3 which means that the grade of relevance of this value is 3.

### 3.7. Building Model

In our study, we used the Tensor Flow-based library for learning-to-rank. This library is considered to be very useful to learn ranking models over massive amounts of data. Tensor Flow is also highly scalable, which contributes to solving the ranking problems in open-source deep learning packages. The proposed deep learning approach is tested in the next section. First, we implement our baseline architecture, and then we enhance its configuration by changing the hyper-parameters and the learning to rank approach (*point-wise*, *pair-wise*, *list-wise*). The aim of these changes is to maximize the ranking related metric and minimize the loss value. The principal goal of our approach is to build a predictive model able to re-rank the best relevant solutions for programming error or bug report using the data from stack overflow. In our framework, the implementation of this problem was inspired from the architecture presented by Pasumarthi [13]. Figure 3 shows the details of our re-rank predictive model.

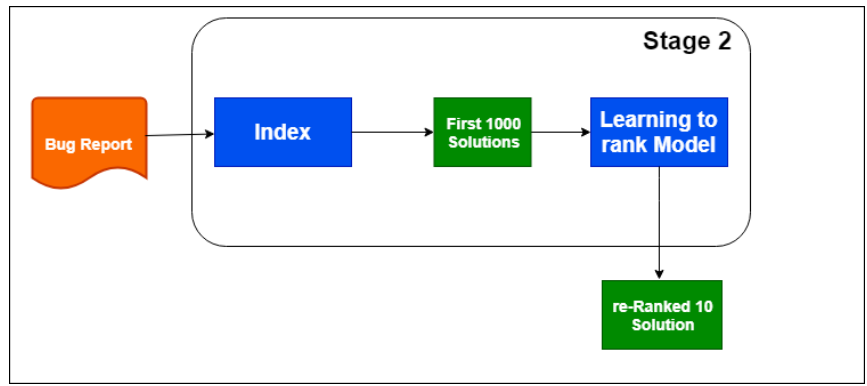

*Figure 3* *Details of our re-rank predictive model*

### 3.8. Evaluation metrics

In this sub-section, we will discuss how we can evaluate the results obtained by our model. We will try to answer the following questions: How do we evaluate the superiority of our model over others? Is there a way to predict that the proposed model will have high quality results in the testing and validation phases?

By conducting a literature review in this field, we found that many evaluation metrics have been proposed to evaluate the performance of models. We have also found that in general, many of the ranking metrics are unique and suitable to the task of ranking, which is usually concerned with the presence of few errors in the higher ranked positions. We will clarify this important principle in the definition of the following metrics:

### 3.8.1. Discounted cumulative gain

We can define Discounted Cumulative Gain (DCG) as a measure of ranking quality that measures the usefulness or gain of a document based on its position in the results list. We accumulate the gains from the top of the results list to the bottom, and at the same time, we discount the gains for each result in the lower ranks. When it comes to comparing DCG with other metrics, one of its most important advantage is that it not only takes care of the case of relevance/non-relevance of the document but it also remain valid in the case where the document has a relevance score as a real number, where $rel_i$ is the relevance of the document at index.

$$DCG@k = \sum_{i=1}^{k} \frac{2^{rel_i} - 1}{\log_2(i+1)}, \quad (1)$$

### 3.8.2. Normalized DCG

The problem with the previous metric (DCG) is that queries with small result sets will always have lower DCG scores than queries that return larger result sets and so this will result in the DCG rising proportionally with $k$ or remaining the same. To solve this problem we usually resort to using a method to compare queries by *normalizing* the DCG score to the maximum at each threshold $k$ in order to ensure that the process is fairer.

$$NDCG@k = \frac{DCG@k}{IDCG@k}, \quad (2)$$

Where IDCG $@k$ represents the best value of DCG $@k$ which can be a DCG value that enables us to reach the best possible ordering of related documents $(rd)$ at threshold $k$, i.e.:

$$IDCG@k = \sum_{i=1}^{rd \ at \ k} \frac{2^{rel_i} - 1}{\log_2(i+1)} \quad (3)$$

### 3.8.3. Average Relevant Precision

We can define Average Relevant Precision (ARP) as a metric that inform us about the degree of concentration of the relevant documents in the highest ranked predictions.

$$ARP = \sum_k (Recall@k - Recall@k - 1) . Precisision@k \quad (4)$$
$$Pecision@k = \frac{true\ positives@k}{(true\ positives@k + false\ positives@k)} \quad (5)$$
$$Recall@k = \frac{true\ positives@k}{true\ positives@k + false\ negatives@k} \quad (6)$$

### 3.8.4. Mean Average Precision

Mean Average Precision (MAP) is metric used when there is a need to know the performance of a model's rankings when evaluated on a whole validation set. This means that MAP is simply an average of ARP over all examples in a validation set.

### 3.8.5. The Mean Reciprocal Rank

We use the Average Reciprocal Rate (MRR) metric when we are caring about having the highest-ranked relevant item. The value of reciprocal rank RR is $\frac{1}{rank}$ where in the case of a single query where the rank is the position of the highest ranked answer out of N answers returned by that query. If no correct answer was returned for the query, then the reciprocal rank is null. In the case of multiple queries $Q$, the MRR is the mean of the $Q$ reciprocal ranks calculated according to the following equation:

$$MRR = \frac{1}{Q} \sum_{i=1}^{Q} \frac{1}{rank_i} \quad (7)$$

## 4. Experiments and results

In this section of the paper, we will address the most important part of our study related to the experiments and discussions. After proposing our model architecture in the previous section, a set of experiments were used with full explanations and details to ascertain the validity of our proposed architecture. We also conducted a series of comparisons between experiments, which are illustrated using tables and graphs.

### 4.1. Experiment Setup

We have conducted five (5) experiments on our dataset. We split our data set into 80% for training and 20% testing. To achieve the goal of not duplicating test set examples in the training set, we have used the principle of splitting so that only multiple rows from the same source are grouped and included in one of the split sets. This will ensure that data from that source is not leaked across multiple sets. This process is an important step to consider when dividing the dataset into subsets for training and testing where there are multiple rows from the same source. For the aforementioned reason and to avoid group bias, the posts by the same users were kept together in either the testing set or the training set during the partitioning.

We used the function *sklearn.model_selection.GroupShuffleSplit* built on a third-party provided group to obtain randomized train/test indices for splitting the data. The group information will be used to encode the specific stratifications of an arbitrary domain of samples as integers. For example, to allow the cross-validation against time-based splits, these groups could be the year of collection of the samples.

The training part contains 3969 queries and 19675 answers. The testing collection contains 974 queries and 4777 answers. The total size of vocabulary is 62272 tokens. The details are shown in Table 4.

|  | Training | Testing | Total |
|---|---|---|---|
| **Query** | 3969 | 974 | 4943 |
| **Answer** | 19675 | 4777 | 24452 |
| **Total** | 23644 | 5751 | 29395 |
| **Total number of tokens** |  |  | 62272 |

*Table 4 Details of our dataset.*

We opted for the training of our model to use the *Google Colab's* platform to speed up the training process. The goal of our experiments is to maximize the ranking metrics by changing the hyper-parameters of our deep learning learn-to-rank architecture (*pointwise, pair-wise, list-wise*). To ensure the saving of our model's results, we used a Model checkpoint. This monitor parameter allows us to save the model state every 1000 steps.

In this section a set of experiments will be conducted with an analysis of their results to obtain a clearer picture of the performance of our approach.

### 4.2. Research Questions

Our analysis is primarily based on examining the results for each experiment separately, and then comparing the overall results among all experiments. In order to reach our goals from this study, we identified five (5) basic research questions, which are as follows:

- RQ-1: *How effective is our model with the use of textual/embedding features of question-answer pairs?*
- RQ-2: *What is the effectiveness of our model under different parameter settings?*
- RQ-3: *How effective is our model with the use of dense (textual, community and affective) features?*
- RQ-4: *What is the effectiveness of our model under a qualitative evaluation?*
- RQ-5: *How effective is our model compared against the state-of-the-art models?*

### 4.3. Baselines

In order to investigate the effect of different components on our model, we did compare our proposed learn-to-rank model with baseline models. The baselines were designed by tuning the parameters of certain components. We can look to these baselines as simplified variations of our model as follow:

• **Our Baseline 1 (D=3,TE,LR=0.001)**, which is our model with a neural network composed of 3 *dense* layers trained on *Textual / Embedding Features* with application of a *learning rate* equal to 0.001.

• **Our Baseline 2 (D=4, TE, LR=0.05)**, which is our model with a neural network composed of 4 *dense* layers trained on *Textual / Embedding Features* with application of a *learning rate* equal to 0.05.

### 4.4. Experiment I (RQ-1: Learn-to-Rank + Textual / Embedding Features)

In this experiment, we will provide the experimental results to the first investigated research question RQ-1 raised in Section 4.2. We proposed a deep artificial neural network architecture consisting of 3 dense layers [64, 32, 16] with *point-wise* approach *and Relu* activation function. The hyper-parameters of this architecture are described in Table 5.

| Parameter | value |
|---|---|
| Batch size | 32 |
| Number of epochs | 15*1000 |
| Optimizer | Adagrad |
| Learning rate | 0.001 |
| Dropout rate | / |
| LIST SIZE | 50 |
| Groupe Size | 1 |
| Loss function | APPROX_NDCG_LOSS |
| global_step/sec | 33.8069 |

*Table 5 First Experiment Hyper-parametres*

The results of this experiment are presented in Table 6. To provide a better visualization of the results we have also plotted them in Figure 4.

|  | @1 | @2 | @3 | @4 | @5 | @6 | @7 | @8 | @9 | @10 |
|---|---|---|---|---|---|---|---|---|---|---|
| **NDCG** | 0.49 | 0.60 | 0.67 | 0.72 | 0.75 | 0.76 | 0.77 | 0.78 | 0.78 | 0.78 |
| **Precision** | 1 | 0.95 | 0.88 | 0.80 | 0.72 | 0.65 | 0.59 | 0.53 | 0.48 | 0.44 |
| **ARP** | 4.64 | | | | | | | | | |
| **MRR** | 1 | | | | | | | | | |

*Table 6 First Architecture metric results*

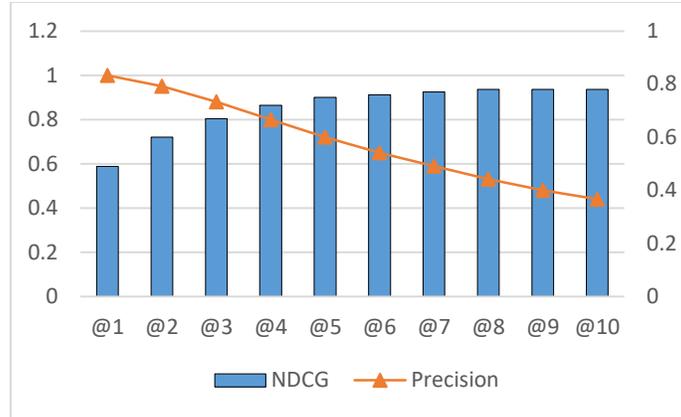

*Figure 4* Performances of Experiment I

*Results Discussion for Experiment I*

Figure 4 above is indicating to what extent our model is not overfitting over the training data. We can clearly see how accuracy as well as loss curves have a monotonic decrease throughout the training process.

From this experiment, we obtained an approximated NDCG loss of **-0.79**. We obtained an NDCG of 49 % for the first element, 75% for the fifth and 78% for the element @10. The precision was of 100% for the first element, 72% for the fifth and 44% for the last element. The ARP was 4.64 and the MRR was 100%. From this experiment, we can conclude that for our algorithm, P@N decreases with N and NDCG@N tends to stay constant with N. In the next experiment, we will add some regularization to our model and see if the NDCG metric and the loss will be enhanced.

**4.5. Experiment II (RQ-2: Learn-to-Rank + T/E Features under Different Parameters Setting)**

In this experiment, we add another layer to the first architecture in order to investigate the second research question RQ-2 and provide appropriate experimental results. We increase the number of neurons in this layer by four ascendant values [32, 64, 128, and 512]. We also increased the dropout and learning rate values. Table 7 illustrate the hyper-parameters used to build our model. Table 8 shows the results in detail.

| Parameter | value |
|---|---|
| Batch size | 32 |
| Number of epochs | 15*1000 |
| Optimizer | Adagrad |
| Learning rate | 0.05 |
| Dropout rate | 0.4 |
| LIST SIZE | 20 |
| Groupe Size | 1 |
| Loss function | APPROX_NDCG_LOSS |
| global_step/sec | 33.8069 |

*Table 7* Second Experiment Hyper-parameters

|  | @1 | @2 | @3 | @4 | @5 | @6 | @7 | @8 | @9 | @10 |
|---|---|---|---|---|---|---|---|---|---|---|
| **NDCG** | 0.58 | 0.68 | 0.74 | 0.77 | 0.80 | 0.81 | 0.81 | 0.82 | 0.82 | 0.83 |
| **Precision** | 0.99 | 0.95 | 0.88 | 0.80 | 0.72 | 0.65 | 0.59 | 0.53 | 0.48 | 0.44 |
| **ARP** | 5.12 | | | | | | | | | |
| **MRR** | 0.99 | | | | | | | | | |

*Table 8* Second Architecture metric results

Figure 5 shows the approximated NDCG loss obtained by our model based on the hyper-parameters of experiment II.

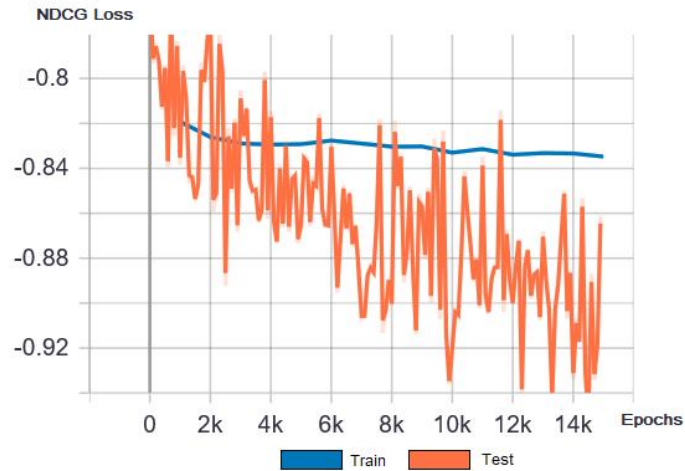

**Figure 5** *Model's Approx NDCG loss obtained by the second architecture*

Figure 6 shows how well the NDCG from top@1 to top @10 is in evolution with each epoch that represent a full iteration on the dataset. The model is improved compared with the first results related to the initial configuration.

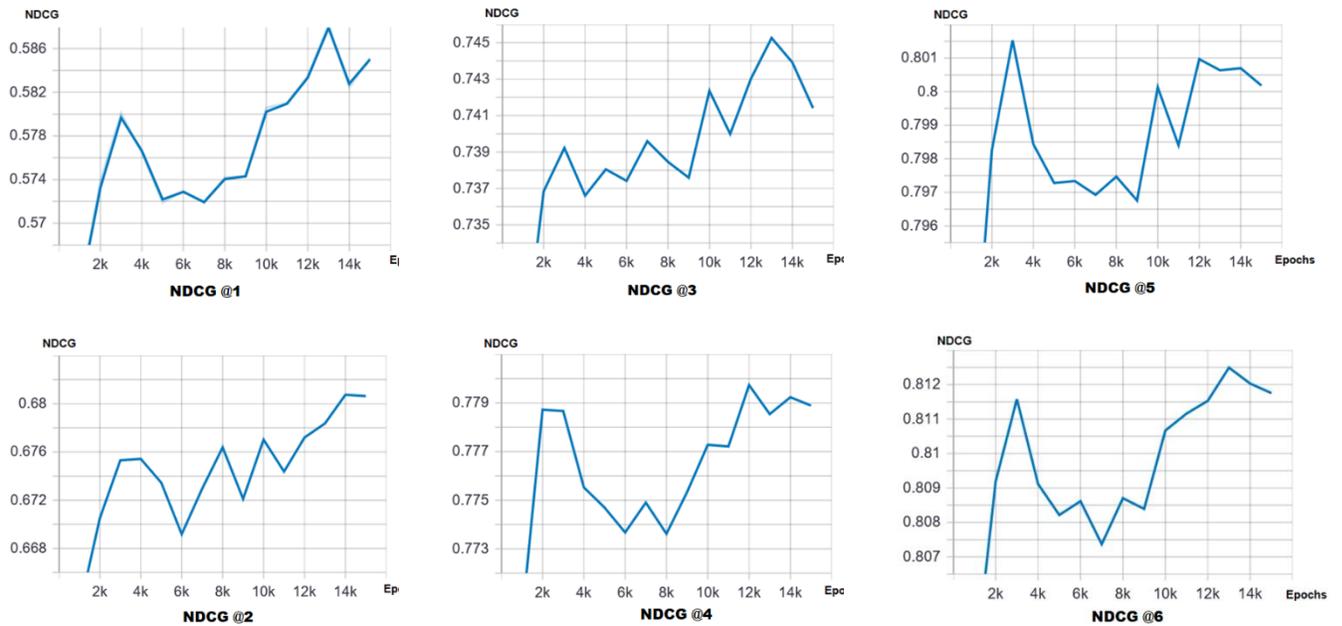

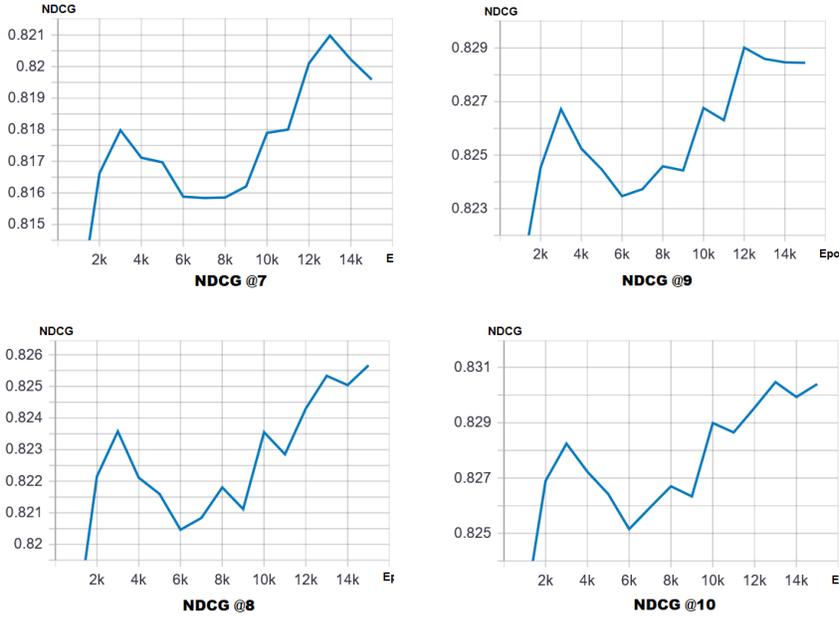

*Figure 6* *Plots of NDCG@10 scores against training epochs obtained by the second architecture during the test stage.*

Figure 7 below show the precision from top@1 to top@10 values in evaluation step. We observed higher value in the first document ranked at 0.99 and low value for the last document ranked at 0.44.

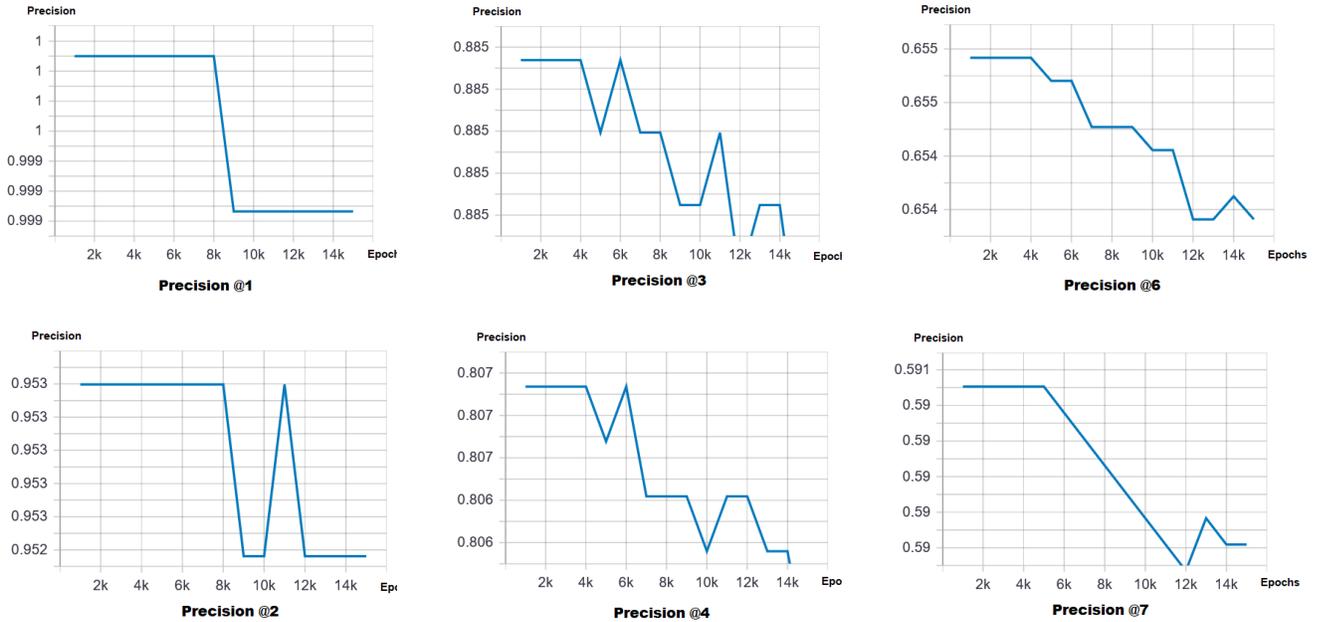

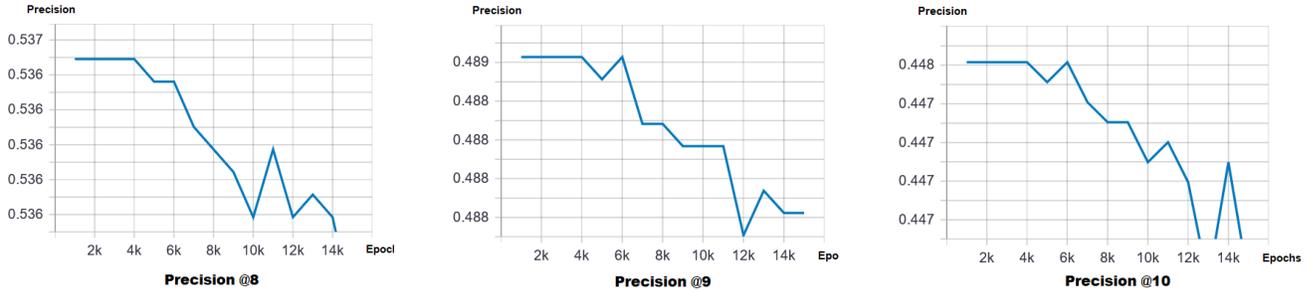

*Figure 7* Plots of Precision@10 scores against training epochs obtained by the second architecture during the test stage.

*Results Discussion for Experiment II*

In the second experiment, we observe that our model has effectively improved. The loss value decrease to -83.4%. Figure 5 presents the extent of change of loss value during the training and testing phases. During the testing phase, the loss function has improved from 75% at 2k epochs to 92% at 14k epochs. The ARP was 5.12 and the MRR was 99%.

Figure 6 shows the plots of NDCG@10 scores against training epochs on all test sets. As seen in the NDCG@N plots, the curve of model reaches its first high value in range of 2k to 4k epochs. The curve of the model is then reduced to the lowest value in the range of 4k to 9k epochs. The second highest value of the curve of our model is reached in the range of 12k to 14k epochs.

Figure 7 shows the plots of Precision@10 scores against training epochs on all test sets. Through these curves, we notice that all of them have a downward trend, which is normal for the precision metric when scored against epochs. With the increasing number of epochs, data can cause some kind of noise over the algorithm, which explain the decreasing of precision values.

As seen in the Precision@N plots, the curve of model reaches its highest value in range of 2k to 4k epochs. The curve of the model is then reduced to the lowest value in the range of 9k to 10k epochs for Precsion@1 to Precison@5. The curve of the model is also reduced to the lowest value in the range of 10k to 13k epochs for Precsion@6 to Precison@10, where a second highest value of the curve of our model is reached in the range of 12k to 13k epochs for Precision@2, Precision@4 and Precision@8.

**4.6. Experiment III: (RQ-3: Learn-to-Rank + Dense Features (Affective, User & Textual Analysis))**

In this experiment, we added dense features to our model in order to obtain the experimental results that will represent the answer to the investigation on the third research question RQ-3. As we said previously, we proceed with the extraction of this kind of features during the preprocessing step. We focus on the following features:

- **Affective features:**
    - Sentiment
    - Subjectivity,
- **User features:**
    - Reputation
- **Textual analysis features**
    - Body length,
    - Title body similarity,
    - Code percentage and
    - Index of readability.

Table 9 presents the results of this experiment.

|  | @1 | @2 | @3 | @4 | @5 | @6 | @7 | @8 | @9 | @10 |
|---|---|---|---|---|---|---|---|---|---|---|
| NDCG | 0.63 | 0.70 | 0.76 | 0.79 | 0.82 | 0.83 | 0.83 | 0.84 | 0.84 | 0.84 |
| Precision | 1.0 | 0.95 | 0.88 | 0.80 | 0.72 | 0.65 | 0.59 | 0.53 | 0.48 | 0.44 |
| ARP | 5.3 | | | | | | | | | |
| MRR | 0.98 | | | | | | | | | |

*Table 9 Third Architecture metric results*

Figure 8 below is showing the approximated NDCG loss obtained by our model based on the hyper-parameters of experiment III.

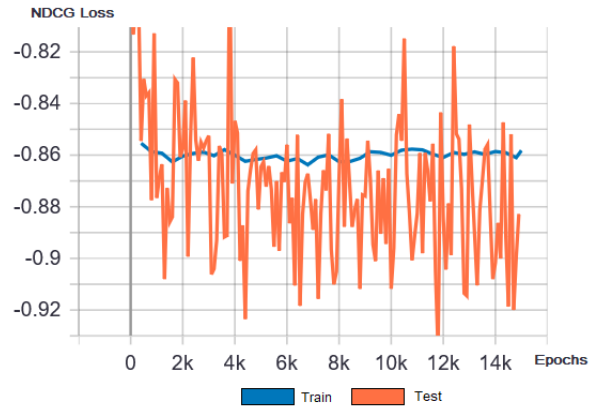

*Figure 8 Model's Approx NDCG loss obtained by the third architecture*

Figure 9 shows how well the NDCG of top @1, top@2 and top@10 is in evolution with each epoch. The model is improved compared with the first results related to the initial configuration.

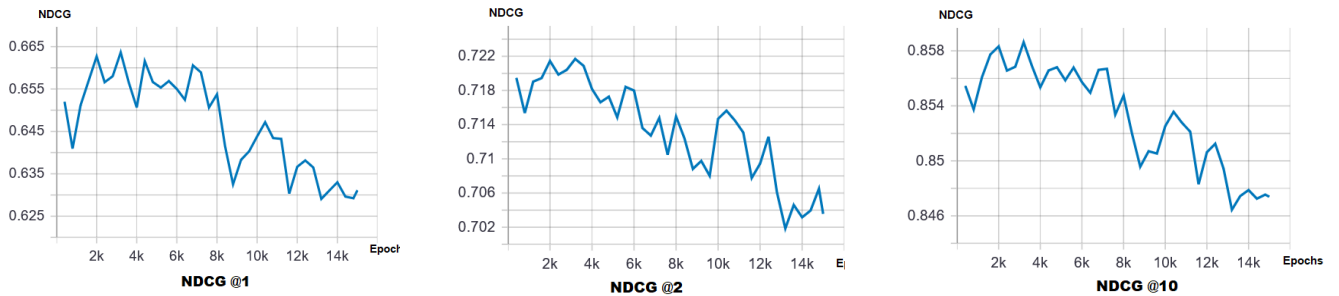

*Figure 9 Plots of NDCG@10 scores against training epochs obtained by the third architecture during the test stage*

The Figure 10 below show the precision@1, precision@2 and precision@10 values in evaluation step We observed higher value in the first document ranked at 0.99 and low value for the last document ranked at 0.44.

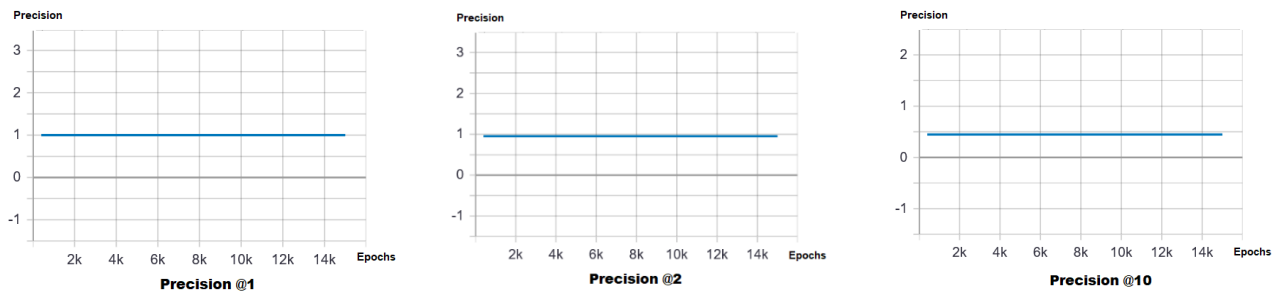

**Figure 10** *Plots of Precision@10 scores against training epochs obtained by the third architecture during the test stage.*

*Results Discussion for Experiment III*

In the third experiment, we observe that our model is effectively improved. The loss value decrease to -86%. The change of loss value during the training and testing epochs is showed in Figure 8 above. During the testing phase, the loss function is improved from the value 82% at 0-2k epochs until 93% at 12-14k epochs. The ARP was 5.30 and the MRR was 98%.

Figure 9 shows the plots of NDCG@1, NDCG@2 and NDCG@10 scores against training epochs on all test sets. For the three plots, the curve of model reaches the highest value in the range of 2k to 4k epochs and the lowest value in the range of 12k to 14k epochs. The second highest value of the three curves have a curve of our model is reached in the range of 12k to 14k epochs. We also notice that all the curves have a downward trend.

Figure 10 shows the plots of Precision@1, Precision@2 and Precision@10 scores against training epochs on all test sets. Through these curves, we notice that for our algorithm, P@N tends to stay constant with N. This confirms that the improvements that we made by adding dense features to the model allowed us to reach much better results. With the increasing number of epochs, adding more data does not influence the convergence of our model, which explain the stability of precision values.

### 4.7. Comparison and discussion of all experiments results

Our approach makes use of the Learn-to-Rank (LTR) technique. To better understand the improvement in the performance of our model, we performed a comparative analysis in which we could identify the individual contribution of each experiment to performance. We gradually increased the size of the training data set over all experiments. We calculated the results for testing data for each experiment and then the average (over all stages) of the LTR metrics (NDCG@N, Precision@N, ARP, and MMR).

|  |  | @1 | @2 | @3 | @4 | @5 | @6 | @7 | @8 | @9 | @1 |
|---|---|---|---|---|---|---|---|---|---|---|---|
| **LTR + T/E Features + (LR=0.01;DO=/;LS=50)** | **NDCG** | 0.4 | 0.6 | 0.6 | 0.7 | 0.7 | 0.7 | 0.7 | 0.7 | 0.7 | 0.78 |
|  | **Precision** | **1.0** | 0.9 | 0.8 | 0.8 | 0.7 | 0.6 | 0.5 | 0.5 | 0.4 | 0.44 |
|  | **ARP** | 4.64 | | | | | | | | | |
|  | **MRR** | 1 | | | | | | | | | |
| **LTR + T/E Features + (LR=0.05;DO=0.4;LS=20)** | **NDCG** | 0.5 | 0.6 | 0.7 | 0.7 | 0.8 | 0.8 | 0.8 | 0.8 | 0.8 | 0.83 |
|  | **Precision** | 0.9 | 0.9 | 0.8 | 0.8 | 0.7 | 0.6 | 0.5 | 0.5 | 0.4 | 0.44 |
|  | **ARP** | 5.12 | | | | | | | | | |
|  | **MRR** | 0.99 | | | | | | | | | |
| **LTR + (T/E;D) Features + (LR=0.05;DO=0.4;LS=20)** | **NDCG** | 0.6 | 0.7 | 0.7 | 0.7 | 0.8 | 0.8 | 0.8 | 0.8 | 0.8 | **0.84** |
|  | **Precision** | **1.0** | 0.9 | 0.8 | 0.8 | 0.7 | 0.6 | 0.5 | 0.5 | 0.4 | 0.44 |
|  | **ARP** | **5.3** | | | | | | | | | |
|  | **MRR** | 0.98 | | | | | | | | | |

*Table 11 Combined results of the three experiments.*

The experimental results are given in Table 10 from which, we can see the role that learning-to-rank technique play in the improvement of answers recommendation. The three experiments make different contributions using the same training and testing datasets. Experiment III performs better than experiment II and Experiment I in terms of NDCG, Precision and ARP. Experiment I is the lowest among the three Experiments, but gets the highest MRR with our approach.

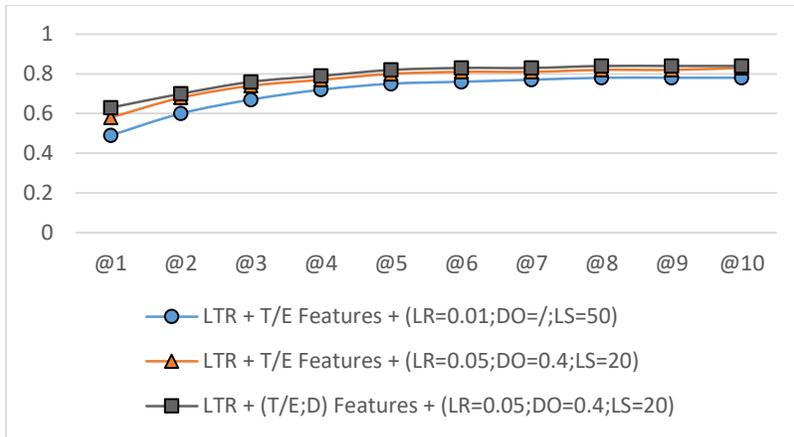

*Figure 11* Plots of NDCG @10 scores obtained by the three experiments

To give a better visualization of the obtained results, we have also plotted the results in the figures Fig. 11, Fig. 12 and Fig. 13. From these figures, we can notice that the performance improves with the adjustments we made on the configuration of our model. This observation was validated across all the three experiments, indicating the feasibility of our approach for bug solutions-based recommendation. More specifically, the proposed model has considerably enhanced all evaluation metrics. The ARP was improved by 12.45%, while the MMR has decreased by 2%.

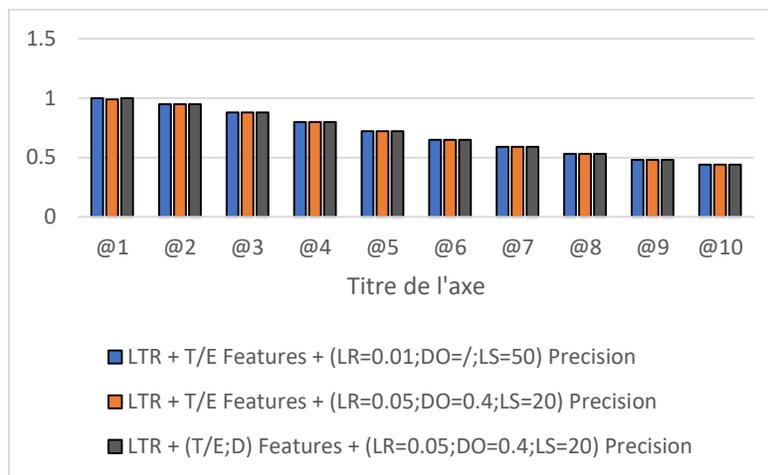

*Figure 12* Plots of Precision @10 scores obtained by the three experiments

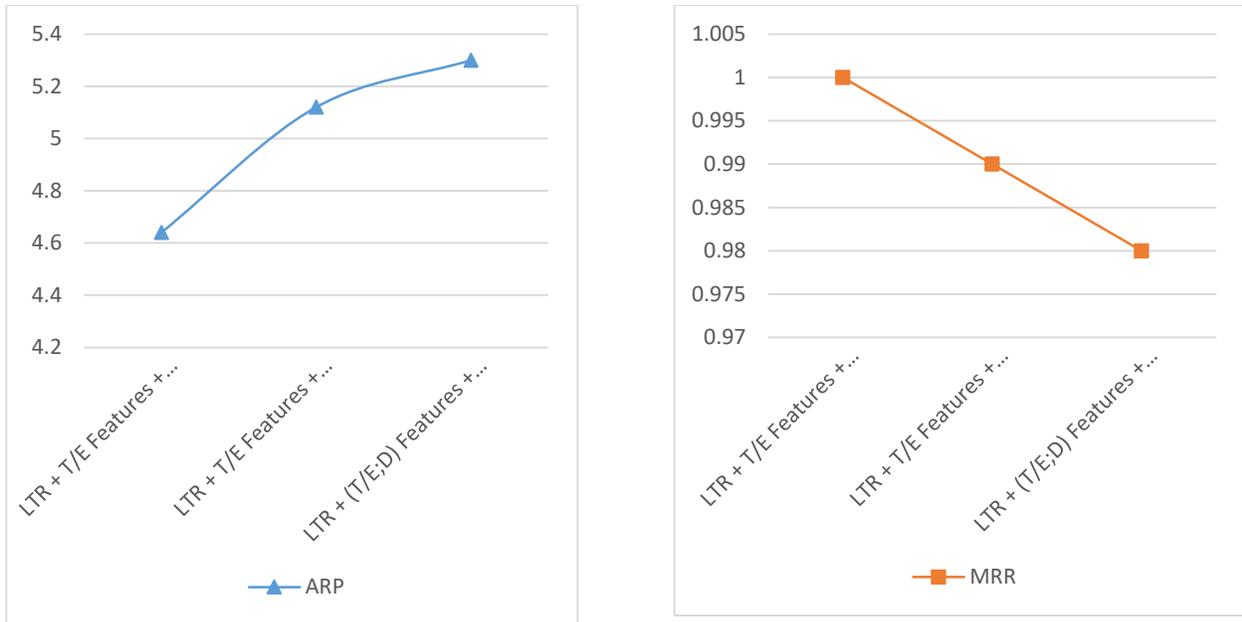

*Figure 13* Plots of ARP @10 & ARP@10 scores obtained by the three experiments

Comparing the results between the three experiments at NDCG@10 and Precision@10, we can see from Fig. 11 that our model is achieving the best results on the whole when setting (Learning Rate=0.05; Drop Out=0.4; List Size=20), especially when recommending 10 Answers. For example, our model achieves Precision@10 and NDCG@10 of 44% and 84% for Experiment III. In contrast, when setting when setting (Learning Rate=0.01; Drop Out=0.0; List Size=50), our model achieves 44% in terms of Precision@10 and 78% in terms of NDCG@10, respectively. As for MAP and MRR, our model also achieves the best results when setting (Learning Rate=0.05; Drop Out=0.4; List Size=20).

### 4.8. Experiment IV: Qualitative Evaluation (RQ4: To what extent the provided answers can really help developers?)

In this experiment we will provide the experimental results of the investigation of the fourth research question RQ-4 raised in Section 4.2.

**Motivation:** We are interested in investigating the question of how the actual returned Stack overflow posts can be used to fix a given bug. We are also interested to know if we can find those posts into the top N ranked answers after similarity analysis done by our system. Knowing this will help Stack Overflow plan better ways to propose the best relevant answers. We can also provide insights for users to give more importance to such answers.

**Approach:** We perform a limited scale qualitative analysis to understand to what extent the provided answers can really help developers. In this experiment, we randomly selected two queries Q1={What causes a java Array Index Out Of Bounds Exception and how do I prevent it?} and Q2={How do you connect to a MySQL database in Java using MySQL JDBC Driver?} (including all their associated answers) out of the 313 answers of Q1 and of the 217 answers of Q2. The results of this study are shown in Table 1.

In this important part of the research, we conducted a qualitative study by evaluating the user experience of our proposed system with two Java language developers (E1, E2). We instructed participants in our system's judging to report relevant or irrelevant posts for each query. E1 and E2 then discussed the answers provided by our system to consider disagreements in order to reach consensus. Table 12 presents the evaluation results of Q1. These results are better seen in Figures 14 and Figure 15.

| Answers | Votes | E1 | P@K | R@K | E2 | Pr@K | R@K |
|---|---|---|---|---|---|---|---|
| A1 | 313 | R | 1.00 | 0.08 | R | 1.00 | 0.08 |
| A2 | 55 | R | 1.00 | 0.15 | R | 1.00 | 0.15 |
| A3 | 24 | NR | 0.67 | 0.15 | NR | 0.67 | 0.15 |
| A4 | 18 | R | 0.75 | 0.23 | R | 0.75 | 0.23 |
| A5 | 13 | R | 0.80 | 0.31 | R | 0.80 | 0.31 |
| A6 | 12 | R | 0.83 | 0.38 | R | 0.83 | 0.38 |
| A7 | 7 | R | 0.86 | 0.46 | NR | 0.71 | 0.38 |
| A8 | 3 | R | 0.88 | 0.54 | NR | 0.63 | 0.38 |
| A9 | 1 | NR | 0.78 | 0.54 | NR | 0.56 | 0.38 |
| A10 | 1 | NR | 0.70 | 0.54 | NR | 0.50 | 0.38 |

**Table 12** *Evaluation metric results of Q1.*

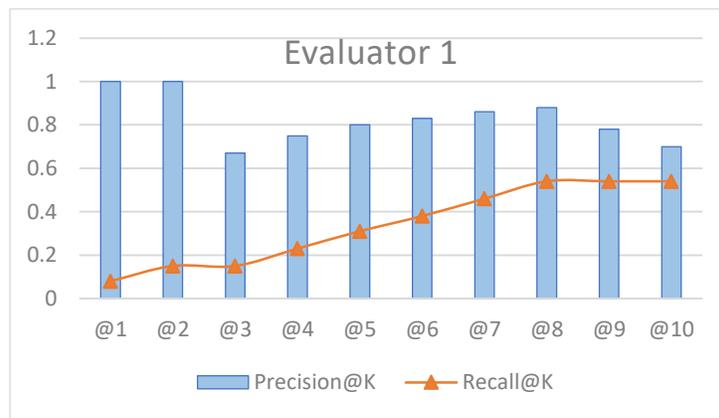

*Figure 14 Performances of Evaluator 1 for Q1.*

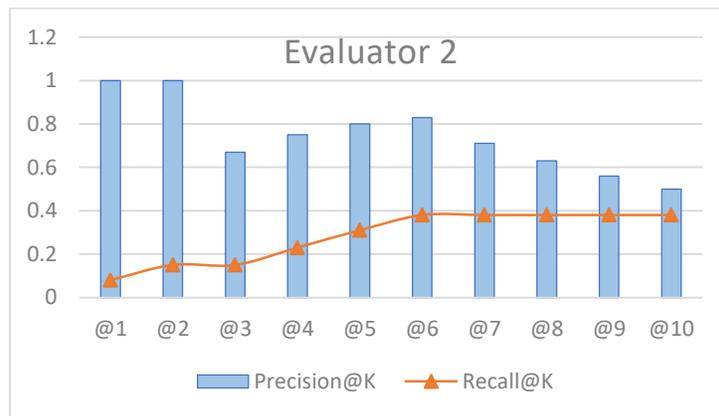

*Figure 15 Performances of Evaluator 2 for Q1.*

Table 13 with Figure 16 and Figure 17 are presenting the evaluation results of Q2.

| Answers | Vote | E1 | P@K | R@K | E2 | P@K | R@K |
|---|---|---|---|---|---|---|---|
| A1 | 217 | R | 1.00 | 0.10 | R | 1.00 | 0.10 |
| A2 | 496 | R | 1.00 | 0.20 | R | 1.00 | 0.20 |
| A3 | 43 | R | 1.00 | 0.30 | R | 1.00 | 0.30 |
| A4 | 27 | NR | 0.75 | 0.30 | R | 1.00 | 0.40 |
| A5 | 12 | R | 0.80 | 0.40 | NR | 0.80 | 0.40 |
| A6 | 2 | R | 0.83 | 0.50 | R | 0.83 | 0.50 |
| A7 | 2 | R | 0.86 | 0.60 | R | 0.86 | 0.60 |
| A8 | 1 | NR | 0.75 | 0.60 | R | 0.87 | 0.70 |
| A9 | -1 | NR | 0.67 | 0.60 | NR | 0.78 | 0.70 |
| A10 | -2 | NR | 0.60 | 0.60 | NR | 0.70 | 0.70 |

**Table 13** *Evaluation metric results of Q2.*

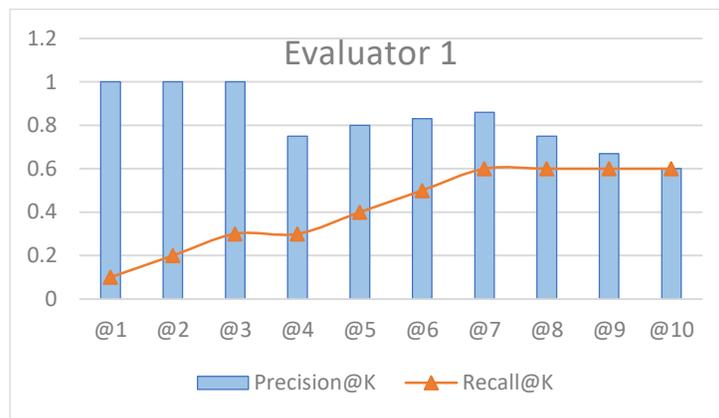

**Figure 16** *Performances of Evaluator 1 for Q2.*

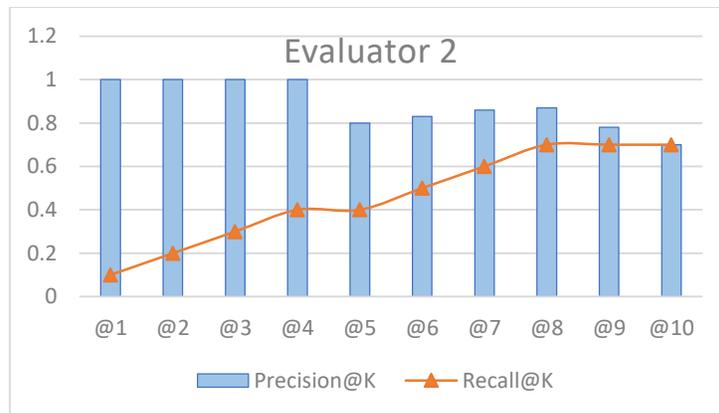

**Figure 17** *Performances of Evaluator 2 for Q2.*

Tables 12 and 13 shows the evaluations made by the two evaluators on the set of solutions compared to the solutions obtained by our system. To better understand the general trends of the human qualitative assessment, we used Figures 14, 15, 16, and 17 to show where and how often judges agree or disagree about the relevancy of a solution. For example for query1, both evaluators marked A1 and A2 as relevant and A3 as non-relevant. The evaluator2 disagrees with evaluator1 and marked the A7 and A8 as non-relevant solutions. The solution where evaluators are in agreement are A1, A2, A3, A4, A5, A6, A9 and A10 for Query 1 and A1, A2, A3, A6, A7, A9 and A10 for Query 2. Evaluators are in disagreement for the following solutions: A7 and A8 for Query 1 and A4, A5 and A8 for Query 2.

*Findings for Experiment IV.*
By referring to the studies conducted in the field of human qualitative evaluation of Questions/Answering systems, we found that many researchers indicated the importance of taking into account agreement among evaluators. Expected and unexpected agreement is inferred by calculating whether the evaluators agree significantly. In this context, many researchers have recommended the use of the Kappa coefficient (K), which allows the measurement of the paired agreements among a group of evaluators making judgment categories. The method of calculating the coefficient is by doing by paired comparisons against an expert or comparing to the decisions of a group of evaluators. The decision is considered to be good if $K > 0.8$, whereas the decision is providing acceptable experimental conclusions if $0.60 < K < 0.8$. In our assessment, we believe that two evaluators agreeing to a decision is an acceptable and appropriate way to consider the right decision about the solution. By calculating the kappa coefficient and comparing our results for "K", it was found that the solutions suggested by our LTR-based system are acceptable. We also found that the inter-raters agreement has a Cohen's kappa value of 0.76 for the Q1 and 0.63 for Q2 which indicates a good agreement between the two evaluators.

.

### 4.9. Experiment V: Comparison of performances (RQ5: Does the proposed model outperform the state-of-the-art models?)

In this experiment, we will provide the experimental results of the investigation of the fifth research question RQ-5 raised in Section 4.2. The aim of this experiment was to study the performance of our proposed model compared to baseline and state of the art models when applied to the same data set. For this purpose, we have accomplished this experiment by proposing an architecture of a deep artificial neural network of 3 dense layer [64, 32, 16] with *Relu* activation function and point-wise approach.

*4.9.1. Comparison with baselines.*

In this section of our fifth experiment, we compared our model with other baseline models. We summarized the evaluations results of this comparison in Table 14. Figure 18 below gives a clearer picture of the results of this comparison.

| Model | NDCG | Precision | ARP | MRR |
|---|---|---|---|---|
| **Our Baseline 1** | 78.00 | 44.00 | 04.64 | 1.00 |
| **Our Baseline 2** | 83.00 | 44.00 | 05.12 | 0.99 |
| **Our Full LTR Model** | 84.00 | 89.90 | 05.30 | 0.98 |

**Table 14** *Performance of three configurations of our system in terms of NDCG@K, Prcision@K, ARP, and MRR, for K=10*

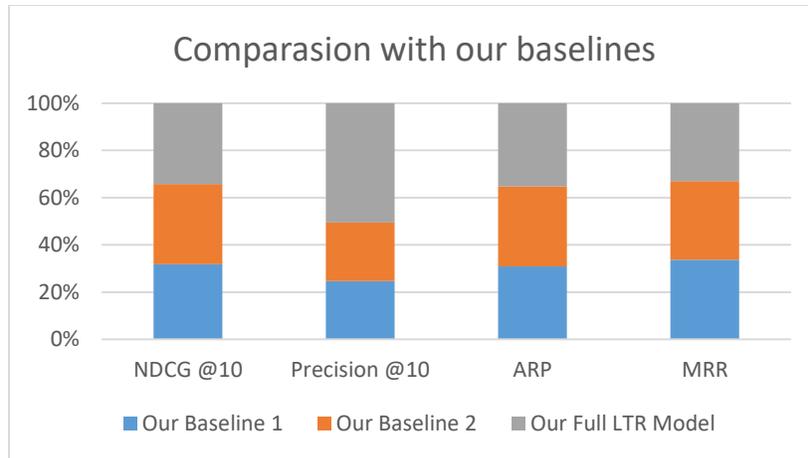

*Figure 18* Our Full LTR model Vs our baselines models.

Table 14 shows the evaluation results of the effectiveness of our proposed model compared to the baseline models when applied to the same dataset. From these results, we can note that the best baseline model results were reported with an NDCG of 83% for baseline model 2 with ($D = 4$, TE, LR = 0.05) as hyperparameters. The results of our proposed model showed its effectiveness compared to the baselines (NDCG = 84% and accuracy = 89.9%).

*4.9.2. Comparison with Search Engines.*

In the second phase of this experiment, a qualitative study was conducted with the aim of understanding how our system would compare to other approaches. We evaluated search results where users try to find solutions in Stack Overflow content throughout Google as the primary search engine or using the integrated Stack Overflow search engine. In this study, results were obtained with Q1="*What causes a java Array Index Out Of Bounds Exception and how do I prevent it?*" and Q2=" *How do you connect to a MySQL database in Java using MySQL JDBC Driver?*" using Google search engine, Stack Overflow search engine in May, 2021. The primary work in this experiment is to evaluate the three research tools based on the criteria of relevance and ranking quality of the top 5 returned results. For this purpose, two evaluators were invited to assess the performance of our system, Google and Stack overflow separately for each query. Table 15 and Figure 19 synthesize the results of the perceptions for query 1. We can observe that our system solutions extracted from Stack Overflow are more relevant than those returned by the two others search engines.

| Query | Google Search Engine | Rel | Stack Overflow Search Engine | Rel | Our proposed System | Rel | P@K |
|---|---|---|---|---|---|---|---|
| Q1 | What causes a java.lang.ArrayIndexOutOfBoundsException | R | What causes a java.lang.ArrayIndexOutOfBoundsException and how do I prevent it? | R | What causes a java.lang.ArrayIndexOutOfBoundsException | R | 1.00 |
| | How to handle Java Array Index Out of Bounds Exception? | R | Why does this simple code throw ArrayIndexOutOfBoundsException? | R | How to avoid ArrayIndexOutOfBoundsException or ... | R | 0.95 |
| | Understanding Array IndexOutofbounds Exception in Java ... | NR | How to .split() a .txt file in Java [duplicate] | NR | What is java.lang.ArrayIndexOutOfBoundsException? - Stack ... | R | 0.88 |
| | What is the ArrayIndexOutOfBounds exception in Java? | NR | Getting Error On Click When Scrolled To Next Page | NR | Array in Java : java.lang.ArrayIndexOutOfBoundsException | R | 0.80 |

| | Why array index out of bounds exception occurs in Java? | PR | ArrayIndexOutOfBounds Exception in an attempt to find the subarray with the largest sum [duplicate] | R | Purposefully avoid ArrayIndexOutOfBoundsExc eption - Stack ... | PR | 0.72 |

**Table 15** *Comparison between Google search engine, Stack Overflow search engine and our proposed system. Rel: the relevancy metric (AR: Absolutely Relevant, NR: Not Relevant, PR: Partially Relevant). P@K: top 5 Precision ratio) for Query 1.*

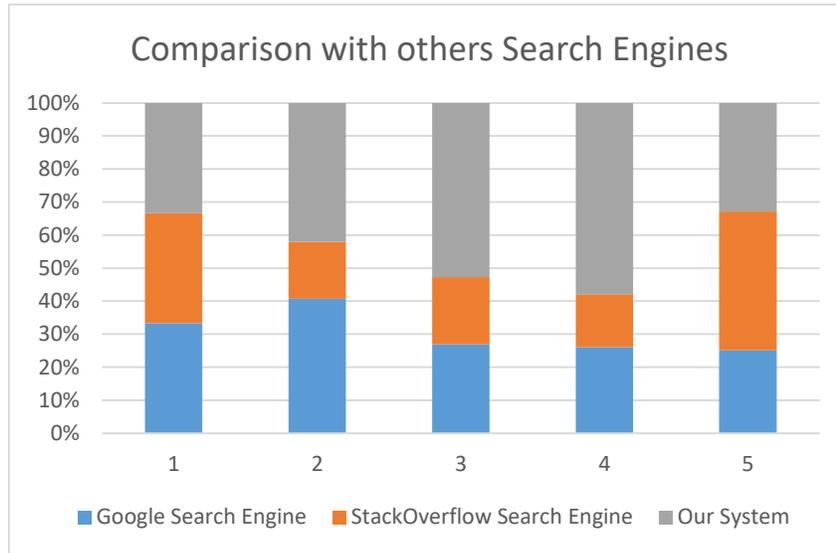

*Figure 19 Results of comparison for Q1.*

Table 16 and Figure 20 synthesize the results of the perceptions for query 2. We can observe that our system solutions extracted from Stack Overflow are more relevant than those returned by the two others search engines.

| Query | Google Search Engine | Rel | Stack Overflow Search Engine | Rel | Our proposed System | Rel | P@K |
|---|---|---|---|---|---|---|---|
| Q2 | Java Database Connectivity with MySQL - javatpoint | NR | java.lang.ClassNotFoundE xception:com.mysql.jdbc.D river [duplicate] | R | getting java.lang.ClassNotFoundEx ception: com.mysql.jdbc ... | R | 1.00 |
| | Connecting to MySQL Using the JDBC DriverManager ... | R | Besides Adding jar for JDBC,"java.lang.classnotf oundexception:com.mysql. jdbc.Driver " | NR | ClassNotFoundException com.mysql.jdbc.Driver - Stack ... | R | 0.90 |
| | MySQL and Java JDBC - Tutorial - vogella.com | NR | Error: java.lang.ClassNotFoundE xception:com.mysql.jdbc.D river [duplicate] | R | java.lang.ClassNotFoundEx ception: com.mysql.jdbc.Driver ... | R | 0.87 |
| | Connect Java to a MySQL database - Stack Overflow | R | JSP & mysql throws java.lang.classNotFoundE xception:com.mysql.jdbc.d river [duplicate] | PR | JAVA-SPRING - Error java.lang.ClassNotFoundEx ception ... | PR | 0.82 |
| | Connecting to MySQL Using JDBC Driver - MySQL Tutorial | PR | Hibernate Reverse Engineering with Eclipse and MySql | NR | Why do I get java.lang.ClassNotFoundEx ception: com.mysql.cj ... | R | 0.75 |

**Table 16** *Comparison between Google search engine, Stack Overflow search engine and our proposed system Rel: the relevancy metric (AR: Absolutely Relevant, NR: Not Relevant, PR: Partially Relevant). P@K: top 5 Precision ratio) for Query 2.*

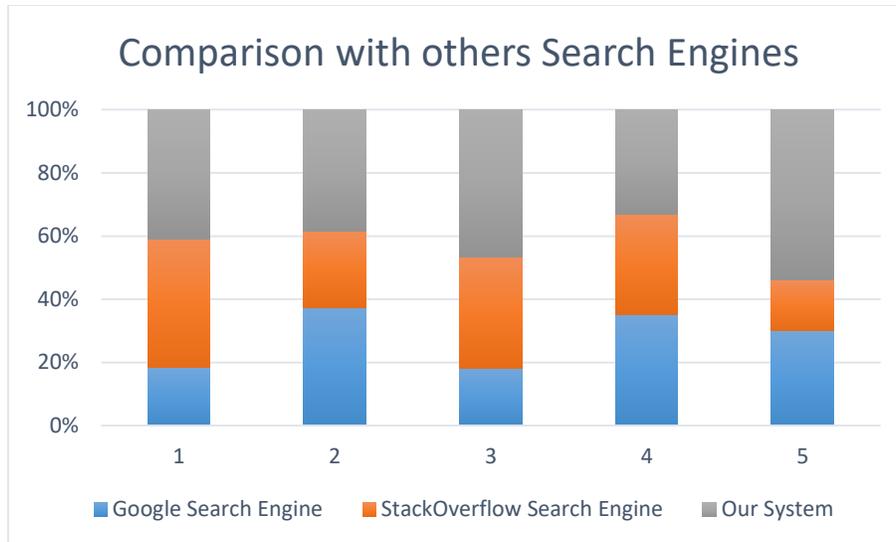

*Figure 20* Results of comparison for Q2.

To measure the performance of the lists of solutions returned by the three compared search engines, we use three relevancy metrics (AR: Absolutely Relevant, NR: Not Relevant and PR: Partially Relevant). Rel represents the degree of relevancy of the solutions in the top 5 obtained results.

To achieve the goal of this experiment, we asked the two evaluators to manually check the three answer lists generated by Google, Stack Overflow, and our proposed system. They tagged the results by marking each solution in the returned list as Absolutely Relevant, Not Relevant or Partially Relevant. The mark indicate the extent to which the evaluator (developer) considered this solution to be relevant to the subject of the query. We asked the evaluators to annotate separately the top 5 solutions for each evaluated query and then to combine their results. The evaluators try to reach a consensus in every case where a mismatch is found while merging the results. The results showed that Cohen's kappa coefficient was equal to 0.733, which indicates that the level of agreement among the evaluators was relatively substantial.

*4.9.3. Comparison with state-of-the-art models.*

The aim of this experiment part is to conduct a comparative study of our proposed model with the most recent models known in this field. Three essential points have been considered in this comparison namely the context of use, the input information and output data used in each model. In this study, we relied on ten recent models shown in Table 17 below. We explain the details of these models as follows:

- The first model, called MAPO, was introduced by Xie et al. [14], and it is about using the existing source code search engines for the purpose of mining API usages from open source repositories. The MAPO system builds a list of method call sequences based on a query consisting of method names or class names.

- Subramanian et al. [15] have developed a system that can correctly link the source code examples to the official API documentation. A constraint-based approach has been adopted in source code snippets to identify fine-grained type references, method calls, and field references.

- Rahman et al. [16] proposed RACK, which, using some code search keywords, can provide a list of relevant API classes as a recommendation to developers. RACK uses regular expressions for building keyword-API associations and extract API class from posts on Stack Overflow.

- Da Silva et al. [17] developed CROKAGE system, which is the acronym of (Crowd Knowledge Answer Generator). This system is based on the idea of taking the description of a programming task (the query) as an input to provide a comprehensible and appropriate solution for the task.

- Li, Xing, & Kabir. [18] proposed CnCxL2R, a software documentation recommendation strategy as a learning-to-rank schema that incorporate the content of official documentation with the social context on Q&A.

- Mahajan et al. [19] proposed Maestro, which a prototype that can automatically assist the developer by recommending the most relevant Stack Overflow posts to a particular Java RE in his code.

- Ponzanelli t, al. [20] proposed a system called Prompter that aims to search for Stack Overflow posts and suggest them in different contexts such as code completion.

- Jiang et al. [21] proposed Deckard, a syntactic code clone detector that is based on comparing ASTs.

- Kisub et al. [22] proposed FaCoY, a system that can detect semantic code clones whereby, by matching the structural code tokens, it can produce a ranked list of Stack Overflow posts.

- McMillan et al. [23] developed a portfolio that can be used for code search task by recommending and visualizing the relevant APIs. The authors demonstrated through their experiments that Portfolio gave an advantage in finding more relevant APIs compared to Google Code Search.

| Model | Context | Query | Response |
|---|---|---|---|
| MAPO [14] | Source Code | API name | API usage |
| Baker [15] | Documentation | Code snippet | Software |
| RACK [16] | Stack Oveflow | Free form | API Class |
| CnCxL2R [17] | Documentation, Stack Oveflow | Free form | Software documentation |
| CROKAGE [18] | Stack Oveflow | Task Description | Code + explanations |
| Maestro [19] | Java RE code | Bug Description | Fixing Runtime Exceptions |
| Prompter [20], | Stack Oveflow | Code snippet | code completion |
| Deckard [21] | Source code | Code snippet | Syntactic code clone |
| FaCoY [22]. | Stack Oveflow | Code snippet | Semantic code clone |
| Portfolio [23] | Source Code | Free form | API usage |
| **Our Full LTR Model** | **Stack Oveflow** | **Free form** | **LTR Solutions** |

**Table 17** *Comparison of our approach with other state of the art models in terms of approaches, context of use, information, input and output data.*

*4.9.4. Findings for Experiment V.*

Tables 15 and 16 show us the performance comparison results of Google search engines and Stack Overflow search engine and our proposed system. We can read the average performance values in terms of the precision in the last row of each table, while the value NR in the "Rel" column indicates that there is no relevant solution returned by the search engine for the query. Our system performed better compared to Stack Overflow search engine and Google search engine with an average Rel value of 4.5 out of 5 solutions.

The results show that for both queries, our proposed system is able to recommend much more relevant solutions at the first position in the list of results, which gives it a performance advantage. Conversely, Stack

Overflow and Google are able to return 3 out of 5 and 2.5 out of 5 solutions relevant to the given query respectively. Through this comparison, it is clear that the improvement we made to our proposed system gave it a significant advantage compared to the Google search engine and the Stack Overflow search engine.

By analyzing the results of this study, we noticed that our proposed system can successfully bridge the lexical gap and identify semantics between query and solutions by capturing the relevance of queries and answers in a lower dimensional semantic space. However, it was not able to accurately recommend good bug documentation or definitions for some queries.

Table 17 is showing the results of the comparison study of our approach with other state-of-the-art models in terms of context of use, input information, and output data. Through this comparison, we noticed the existence of a set of different works from the related works in the context of program analysis and mining code repositories. At the end of this study, we can confirm that our proposed model is a novel bug solutions recommendation approach incorporating the content and social context developed by users in Stack Overflow into one unified learning-to-rank schema.

### 4.10. THREATS TO VALIDITY

This section introduces three different kind of threats to validity. We will discuss the major issues related to threats to internal validity, threats to external validity, and threats to development validity.

*4.10.1. Threats to internal validity*

Biases and experimental errors are considered among the major sources of threats to internal validity [24]. Data bias is considered as the main threats of this category. In our study, we have conducted 3 experiments on our dataset. We divided the dataset into two collections: training and testing to validate the results that were obtained from the average values during the two phases. The learn-to-Rank (LTR) process were used to mine the knowledge from Stack Overflow answers to recommend the top k answers as solutions for a bug software.

*4.10.2. Threats to external validity*

Threats to external validity are generally related to the problem of generalizing the obtained results to other cases that are contextually different from those used in the experiments [24]. Like many other studies, we cannot confirm that our proposed model will be able to fit with any other recommendation approach, but we think that our experimental results can clearly illustrate the performance improvements. Another point of strengths of our model is that in our experiments, we concentrate on software bugs. Our model was designed to be free from any constraint related to the format of the software bug. The proposed framework does not give any importance to the bug's category since this later is considered as simple textual document and preprocessed using standard NLP techniques. We believe that our model can be adapted to recommend solutions from stack overflow for any type of software bugs.

*4.10.3. Threats to development validity*

With regard to threats to development validity, this forces us to carefully consider the amount of effort required by developers in using a hypothetical tool such as the one proposed in this paper. We can be certain that more attention should be given to the aspect of human-machine interaction in this context. We anticipate that our model can be implemented as a standalone tool based on the idea of web services to make it easier for developers to use this new approach. In the default web interface, we expected the existence of an input field from which developers are allowed to copy and paste their bug queries into the web application. The application can also embed the bug report via a bug report encoder and generates the question titles for the developers, where the input to the system will be a bug report text. Developers can click the "Generate" button to submit their query after the code snippet is entered. When the server receives the query sent by the developer, the bug report text is converted into a query, then the Stack

Overflow database search is started and the system returns the top 10 posts with the links to their respective questions. In this way, developers can use this tool to quickly browse related questions and have a better understanding of the problem in the bug report.

## 5. Conclusion

The work proposed in this paper aims to develop a recommendation system based on the learning-to-rank (LTR) approach and deep learning techniques. The main idea of this research was to investigate the use of data mining on Stack Overflow to automatically suggest relevant solutions that fix software bugs and programming errors. Our main goal was to design a system capable to decrease the time generally spent by developers during the manual efforts for fixing bugs or during the consultation of Q&A websites. We followed a clear scientific methodology in our work, as we started the research by discovering this new area known as Mining Software Repositories (MSR), and that was of course after we conducted an overview of areas related to data mining such as text mining, natural language processing, recommendation systems, and deep learning technologies. We believe this new area of research (MSR) will have an important impact on the future of software development.

In terms of contributions, we proposed an architecture of a recommender system that has a core stage based on the TF-IDF index and a learning-to-rank model based on deep learning to recommend relevant solutions for programing error and software bugs.

We tried to change the approach of learning to rank (from point-wise to pair-wise or list-wise) but we do not get good results because of many reasons. The most important reason is the high dimensionality of our large dataset. We believe that we need more time for preprocessing and for applying deeper NLP techniques to screening more features that will be relevant for our context.

For future work, we will try to test our approach of learning-to-rank using pairwise or list-wise techniques with other parameters. We will also try to improve our model performance using features that are more specific in the training phase. Other deep learning algorithms and techniques like transformers and pre-trained Large Models will also be considered for future testing and comparison purposes.

## 6. ACKNOWLEDGMENT

This work was partially supported by the General Directorate of Scientific Research and Technological Development (GDSRTD), Ministry of Higher Education and Scientific Research, Algeria.